# Knowledge Management in Socio-Economic Development of Municipal Units: Basic Concepts

Maria A. Shishanina[1], Anatoly A. Sidorov[1]
[1]Tomsk State University of Control Systems and Radioelectronics
Tomsk, Russian Federation
mariia.a.shishanina@tusur.ru

*Abstract* — **The article discusses the basic concepts of strategic planning in the Russian Federation, highlights the legal, financial and resource features that act as restrictions in decision making in the field of socio-economic development of municipalities. The analysis concluded that to design an adequate model of socio-economic development of municipalities is a very difficult task, particularly when the traditional approaches are applied. To solve the task, we proposed to use the semantic modeling as well as cognitive maps which are able to point out the set of dependencies that arise between factors having a direct impact on socio-economic development.**

## I. Introduction

The territorial features of the Russian Federation imply non-typical decisions, considered as standard in the world, to be applied in management of social and economic development (SED) of regions and municipal units (MUs). The unique features that exist at the level of municipal units development define the essential role of regional and municipal authorities, which deal with the issues of social and economic development of territories. However, the problem is complicated with the market transformations, which the country faced some years ago, that impacted negatively the readiness of the state and municipal authorities to build an effective region-municipality relationship. This fact is explained by regulatory and legal features, funding and the lack of a comprehensive system able to provide information and analytical support that allows predicting the territorial development. In this regard, the scientific rationale for decision-making in the field of SED of MU is relevant regardless of the development level that the territories have, so the purpose of this work is to describe the approach enabling to improve the effectiveness of management decisions in the planning, forecasting and programming in SED of MU.

## II. Special Management Features for Social-Development of Municipal Units

At the basis of SED for the regions and MUs in accordance with the current legislation (Federal law dated on 28.06.2014 No. 172-FL "about the strategic planning in the Russian Federation") there is the strategic management, which is represented in the basic guiding documents - long-term and intermediate-term programs - aimed at the global development in changing environment [1]. Thus, management in SED of MUs – is a constant process aimed at developing, making and applying the managerial decisions including the situation monitoring measures, approaches to the strategy development as well as implementing plans and programs efficacy evaluation. The analysis of normative- legal documents showed the general view of strategic management that includes the following stages:
- strategic analyses;
- strategic synthesis;
- goal-setting;
- project-based activities;
- strategy mechanisms searching;
- results assessment.

As a result, SED management can be presented as a model (fig. 1), where the key role in planning belongs to the state-based authority, since the first-priority purposes of the federal level (shown with the arrows in fig.1) are projected initially for the larger territorial entities (federal regions), the Russian Federation subjects and only then "go down" to the MUs level.

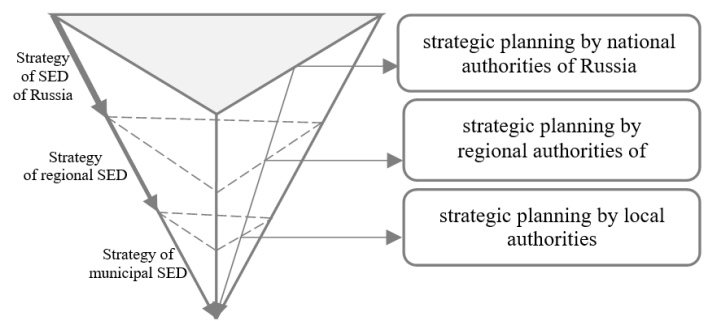

**Figure 1.** "Top-down" SED management

Ideally, the municipal strategic governance as the basis should have strategic plans for local authorities, taking into account the public opinion and the opinion of the business community. However, in practice the situation is different. The local authorities make attempts to implement a strategy based on existing administrative methods. This results in actions only from the part of an executive body, excluding the levels of the local community, business, and etc. Additionally, the staff shortage in MUs impacts negatively, since the existing staff do not have sufficient knowledge in strategic and project management; therefore the changes happening in the external and internal environment aggravate the SED strategy implementation. As a result, the following key challenges in strategic planning at the MUs level can be pointed out [2,3]:

- non-systemic organization of strategic development;
- administrative management methods domination;
- insufficient methodological support from the regional authorities for local initiatives;
- knowledge gap in strategic and project management among officials of different levels (especially in MUs);
- strict financial dependence, since MUs, as a rule, are deprived of additional financial sources;
- lack of an effective mechanism for interaction between local government bodies with the local community, business companies and others parties interested in development.

On the assumption that the strategic planning system at the level of MUs requires rethinking but it is practically impossible to implement at the level of the municipality, excluding the federal level, there is an objective need to increase the level of scientific and methodological substantiation of managerial decisions taken by authorities and administrations in planning and forecasting as well as assessing the effectiveness of socio-economic development of the territory. As a result, the SED management model ("top-down") is transformed into a "bottom-up" model (fig. 2), which will take into account the specifics of the territories when planning long-term development aimed at the national strategic goals.

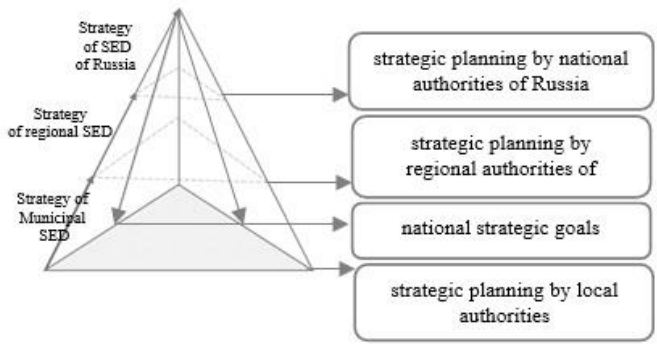

**Figure 2.** Bottom-up SED management

Despite the fact that the model presented in fig. 1 is conceptually different from the model depicted in fig. 2, it can be implemented under current conditions with regulatory, financial, labor and other restrictions. Management of the SED targeted at the territories (especially at MUs) is a non-trivial task, the solution of which requires, inter alia, a creative approach. However, on the part of the legislator, this process is regulated (a list of mandatory documents has been developed, various methods to analyze situations are recommended, etc.), which can actually reduce all creative work to templates use. Still, it is impractical to use one template for all MUs in the Russian Federation (or in a specific region), because each territory is unique and has its own development features.

Except that, it can be assumed that throughout the territory of the Russian Federation there are the MUs with similar levels and vectors in development (for example, climatic conditions, population size, the major industry field, etc.).

As a result, such MUs will act as the standard models, on the basis of which the template-based strategies can be developed. This approach will allow to a certain extent to standardize planning and management processes in SED. As a result, the existing SED system will be based not only on the regional goals, but also on the municipal initiatives, which will vary against the types of territories. Thus, in order to achieve the state goals and increase the effectiveness territories' management at all levels, it is necessary to take into account the peculiarities of MUs development and present the management of SED in a comprehensive manner as a semi-structured system.

### III. APPLICATION OF COGNITIVE TOOLS IN SED MANAGEMENT OF MUS

As noted in [4-6], cognitive modeling of semistructured systems is one of the dimensions of the modern theory of decision support that allow to achieve some adequate results with a large number of interdependent factors. The cognitive modeling has several stages, among which the main one is the identification stage that points out a set of factors and the relationship between them in the semistructured system. The regulatory documents analysis showed that many indicators used for assessing SED model of MUs will act as restrictions to the subject field. Summarizing this set of SED MUs indicators, presented in the strategic documents, the general indicators were identified, on the basis of which a typical cognitive model was designed (fig. 3), which can serve as the basis for modeling SED management on particular territories.

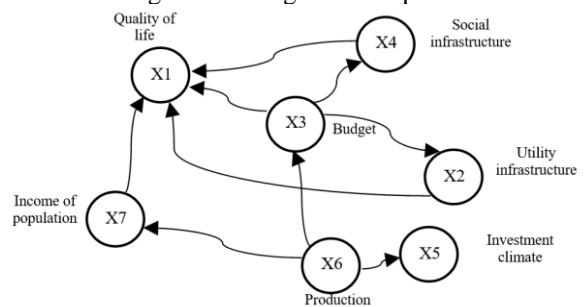

**Figure 3.** Standard cognitive model of SED

Based on this model, a conclusion can be made that the quality of life should be used as the target factor, since federal-level strategic documents identify it as a key factor in municipal SED. "Production" can be used as a variable (special) factor, because it will be showing significant variation depending on the specialization of any given territory (e.g. settlements can be specialized in agriculture, mining, and so on). Each of the factors can be broken down, revealing new levels, links and relationships within the model.

In turn, analysis of various types of cognitive maps [7,8] shows that their tool set is quite adaptive. However, according to the indicated criteria, the Silov's fuzzy cognitive maps are most suitable, since when designing them, it is possible to denote a negative connection and evaluate its impact. It is especially important when considering the SED indicators (for example, crime indicators will have a negative impact on the life quality indicators). Also the fuzzy cognitive maps of Silov have proven themselves well to design the semi-structured systems, which include SED of MUs [9].

In this regard, the fuzzy cognitive map can be represeted as follows: $G = <X, W>$, where $X = \{x1, x2, \ldots, xn\}$ corresponds to a set of concepts (factors) in the subject area (in this case, a set of indicators for evaluation of municipal SED), and $W$ is relations over the set $E$ that determine the combination of connections between elements of the given set. Elements ei and ej are considered connected by the following relation $w(ei,ej) \in W \to [-1, 1]$ if any change in the value of factor ei (cause) results in a change in the value of factor ej (effect). Thus, ei is deemed to have an effect on ej.

The general calculation rule for the values of concepts of Fuzzy Cognitive Map has been proposed [9]:

$$A_i^t = f\left(k_1 \sum_{\substack{j=1 \\ j \neq i}}^{n} A_j^{t-1} W_{ji} + k_2 A_i^{t-1}\right),$$

$A_i^t$ – is the value of concept Xi at time t, $A_j^{t-1}$ – the value of concept Xj at time t-1, $A_i^{t-1}$ – is the value of concept Xi at time t, and the weight $W_{ji}$ – of the interconnection from concept Xj to concept Xi, f is a threshold function which squashes the result in the desired interval. The coefficients $k_1$ and $k_2$ can take different values according to any specific case and are determined by the expert. As a result, the cognitive map allows solving two types of tasks: static (current situation analysis, including the study of effects that some indicators relate to others, the study of situation stability as a whole and the search for structural changes to get the stable structures) and dynamic (generation and analysis of some possible scenarios for evolving situation at different time/over time).

## IV. Conclusion

The SED of the territory is a complex and continuous process of planning and forecasting, which the authority bodies of all levels have to deal with. However, despite the formalized approach to planning at the federal level, municipalities in their activities face a number of challenges. The main difficulties in the decision-making process in the field of SED conclude in the lack of complete information about the territory development from a person who makes decisions. This is due to the fact that SED is the semi-structured subject field and this fact must be taken into account when planning and forecasting are under way. To neutralize the negative impact from the problems, which have been identified in the research, it is proposed to use in management practice the semantic-cognitive tools that will allow to increase the quality and the validity level of managerial decisions made by officials. The proposed approach to formalizing a semi-structured area of public relations can be used as the basis to design the decision-support systems both in municipalities and in other territorial units. This paper is designed as part of the state assignment of the Ministry of Science and Higher Education.

ACKNOWLEDGMENT

This paper is designed as part of the state assignment of the Ministry of Science and Higher Education; project FEWM-2023-0013.